\newcounter{myctr}

\documentclass{ws-acs}
\usepackage{url}
\begin{document}

\makeatletter
\def\@biblabel#1{[#1]}
\makeatother

\markboth{T. Carletti, D. Fanelli and S. Righi}{On the evolution of a social network}

%
\catchline{}{}{}{}{}
%

\title{On the evolution of a social network}

\author{TIMOTEO CARLETTI}

\address{D\'epartement de math\'ematique\\ Facult\'es
  Universitaires Notre Dame de la 
  Paix, rempart de la Vierge 8\\ Namur, B5000, Belgium\\
timoteo.carletti@fundp.ac.be}

\author{DUCCIO FANELLI}

\address{Dipartimento di Energetica and CSDC\\
 Universit\`a di Firenze, and INFN\\
via S. Marta, 3, 50139 Firenze, Italy\\
duccio.fanelli@unifi.it}

\author{SIMONE RIGHI}

\address{D\'epartement de math\'ematique\\ Facult\'es
  Universitaires Notre Dame de la 
  Paix, rempart de la Vierge 8\\ Namur, B5000, Belgium\\
simone.righi@fundp.ac.be}

\maketitle

\begin{history}
\received{(received date)}
\revised{(revised date)}
\end{history}

\begin{abstract}
In this paper we show that the small world and weak ties phenomena can
spontaneously emerge in a social network of interacting agents. This dynamics
is simulated in the framework of a simplified model of opinion diffusion in an
evolving social network where agents are made 
to interact, possibly update their beliefs and modify the social relationships
according to the opinion exchange.  
\end{abstract}


\section{Introduction}
\label{sec:intro}

Modeling social phenomena represents a major challenge that 
has in recent years attracted a growing interest.  Insight into the problem can be gained 
by resorting, among others, to the so called {\em Agent Based Models}, an approach that is well 
suited to bridge the gap between hypotheses 
concerning the microscopic behavior of individual agents and the emergence
of collective phenomena in a population composed of many interacting
heterogeneous entities.

Constructing sound models deputed to return a reasonable approximation of the 
scrutinized dynamics is a delicate operation, given the degree of arbitrariness 
in assigning the rules that govern mutual interactions. 
In the vast majority of cases, data are scarce and do not sufficiently constrain the model, hence the 
provided answers can be questionable. Despite this intrinsic limitation, 
it is however important to inspect the emerging dynamical properties of abstract models, 
formulated so to incorporate the main distinctive traits of a social interaction scheme. 
In this paper we aim at discussing one of such models, by combining analytical and numerical 
techniques. In particular, we will focus on characterizing the evolution of the
underlying social network in terms of dynamical indicators. 

It is nowadays well accepted that several social groups display 
two main features: the {\em small world property}~\cite{WS1998} and the
presence of {\em weak ties}~\cite{Granovetter1983}. The first property implies that the network exhibits clear tendency to organize in large, densely connected, clusters. 
As an example, the probability that two friends of mine are also, and independently, friends 
to each other is large. Moreover, the shortest path between two generic individuals is 
small as compared to the analogous distance computed for a
random network made of the same number of individuals and inter-links connections. This observation signals the  
existence of short cuts in the social tissue. The second property
is related to the cohesion of the group which is mediated by small groups of well tied 
elements, that are conversely weakly connected to other
groups. The skeleton of a social community is hence a hierarchy of subgroups.

A natural question arise on the ubiquity of the aforementioned peculariar aspects, distinctive traits of a real social networks: 
how can they eventually emerge, starting from an finite group of randomly connected actors?
We here provide an answer to this question in the framework of a minimalistic opinion dynamics model, 
which exploit an underlying substrate where opinions can flow. More specifically, 
the network that defines the topological structure is imagined to evolve, coupled to the opinions and following a 
specific set of rules: once two agents reach a compromise and share a common opinion, they also increase their 
mutual degree of acquaintance, so strengthing the reciprocal link. In this respect, the model that we are shortly going to introduce 
hypothesize a co-evolution of opinions and social structure, in the spirit of a genuine adaptive network~\cite{GrossBlasius,ZESM}.

Working within this framework, we will show that an initially generated random group, with
respect to both opinion and social ties, can evolve towards a final state where small worlds and weak ties effects are indeed 
present. The results of this paper constitute the natural follow up of a series of
papers~\cite{pre,opi3,epjb}, where the time evolution of the 
opinions and affinity, together with the  fragmentation vs. polarization 
phenomena, have been discussed. 

Different continuous opinion dynamics models have been presented in
literature, see for instance~\cite{deffuant2000,galam2008}, dealing with the
general consensus  problem. The aim is to shed light onto the assumptions 
that can eventually yield to fixation, a final mono-clustered configuration where all
agents share the same belief, starting from an initial condition where the inspected population 
is instead fragmented into several groups. In doing so, and in most cases, a 
fixed network of interactions is a priori imposed~\cite{AD2004}, and the polarization dynamics studied under the constraint of 
the imposed topology. At variance, and as previously remarked, we will instead allow the underlying network 
to dynamically adjust in time, so modifying its initially imposed characteristics. Let us start by revisiting the 
main ingredients of the model. A more detailed account can be found in ~\cite{pre}.

Consider a closed group of $N$ agents, each one
possessing its own opinion on a given subject. We
here represent the opinion of element $i$ as a continuous real variable
$O_i\in [0,1]$. Each agent is also characterized by its affinity score
with respect to the remaining $N-1$
agents, namely a vector $\alpha_{ij}$, whose entries are real number defined in the interval
$[0,1]$: the larger the value of the affinity $\alpha_{ij}$, the more reliable the relation of $i$ with the end node $j$. 

Both opinion and affinity evolve in time because of binary encounters between
agents. It is likely that more interactions can potentially occur among
individuals that are more affine, as defined by the preceding indicator, or that
share a close opinion on a debated subject. Mathematically, these requirements can be accommodated for by
favoring the encounters between agents that minimizes the {\em social metric}
$D^t_{ij}=|\Delta O_{ij}^t|(1-\alpha_{ij}^t)+\mathcal{N}_j(0,\sigma)$, where
$\Delta O_{ij}^t=O_i^t-O_j^t$ is the opinions' difference of agents $i$ and $j$ at time $t$, and the last term is a stochastic
contribution, normally distributed  with zero mean and
variance $\sigma$. For a more detailed analysis on
the interpretation of $\sigma$ as a {\em social temperature} responsible of a
increased mixing ability of the population, we refer 
to~\cite{pre,opi3,epjb}.

Once two agents are selected for interaction they possibly update their opinions (if they are
affine enough) and/or change their affinities (if they have close enough
opinions), following: 
\begin{equation}
\begin{cases}
O_i^{t+1} &= O_i^{t}- \frac{1}{2} \, \Delta O_{ij}^{t}\,
\Gamma_1\left(\alpha^t_{ij}\right)  \\ 
\alpha_{ij}^{t+1} &= \alpha_{ij}^{t} + \alpha_{ij}^{t}
      (1-\alpha_{ij}^{t}) \,\Gamma_2 \left(\Delta O^t_{ij}\right) \, ,
\end{cases}
\label{eq:themodel}
\end{equation}
being:
\begin{equation}
\Gamma_1 \left(x\right)= \frac{\tanh (\beta_1
  (x-\alpha_c)) + 1 }{2} \quad\text{and}\quad  
\Gamma_2 \left(x\right)= -\tanh(\beta_2 (|x| -
\Delta O_c)) \, ,
\end{equation}
two {\em activating functions} which formally reduce to step functions for large enough the values of
the parameters $\beta_1$ and $\beta_2$, as it is the case in the numerical simulations reported below.

Despite its simplicity the model exhibits an highly non linear dependence on
the involved parameters, $\alpha_c$, $\Delta O_c$ and $\sigma$, with a phase
transition between a polarized and fragmented dynamics~\cite{pre}. 

\begin{figure}[ph]
\centerline{\psfig{file=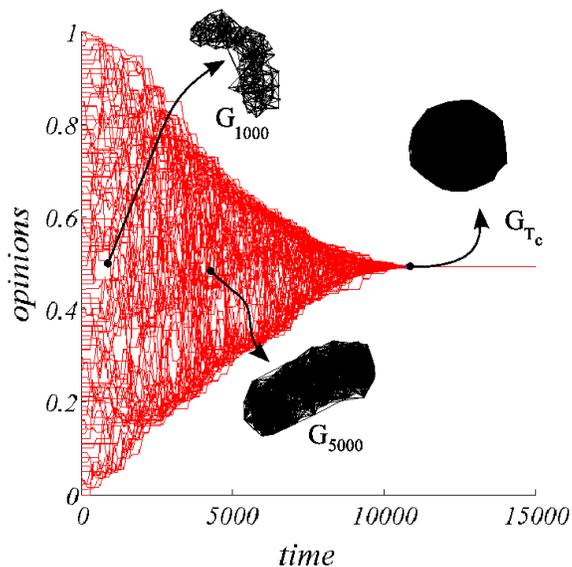,width=8cm}}
\vspace*{8pt}
\caption{Opinions as function of time. The run refers to
 $\alpha_c=0.5$, $\Delta O_c=0.5$ and $\sigma =
  0.01$. The underlying network is displayed at 
  different times, testifying on its natural tendency to evolve towards
  a single cluster of affine individuals. Initial opinions are
  uniformly distributed in the  
interval $[0, 1]$, while $\alpha_{ij}^0$ are randomly assigned in $[0,1/2]$
with uniform distribution.}
\label{fig:fig1}
\end{figure}

A typical run for $N=100$ agents is reported in
the main panel of Fig.~\ref{fig:fig1}, for a choice of the parameters which
yields to a consensus state. The insets represent three successive time snapshots of
the underlying social network: The $N$ nodes are the individuals, while
the links are assigned based on the associated values of the affinity. The figures  
respectively refer to a relatively early stage of the evolution $t=1000$, to an 
intermediate time $t=5000$ and to the convergence time $T_c=10763$. Time is here calculated as the number 
of iterations (not normalized with respect to $N$). The 
corresponding three networks can be characterized using standard topological
indicators~\cite{AB2002,blmch2006} (see Table~\ref{tab:table1}), e.g. the mean
degree $<k>$, the network clustering coefficient $C$ and the average shortest path $<\ell >$. An 
explicit definition of those quantities will be given below.


In the forthcoming discussion we will focus on the evolution of the network
topology, limited to a choice of the parameters that yield to a final mono cluster.

\begin{table}[ph]
\tbl{Topological indicators of the social networks presented in
  Fig.~\ref{fig:fig1}. The mean degree $<k>$, the network clustering $C$ and
  the average shortest path $<\ell >$ are 
  reported for the three time configurations depicted in the figure.}
{\begin{tabular}{@{}cccc@{}} \toprule
 &  $<k>(t)$ & $C(t)$ & $<\ell >(t)$ \\ \colrule
$t=1000$ & $0.073$ & $0.120$ & $3.292$ \\
$t=5000$ & $0.244$ & $0.337$ & $2.013$ \\
$t=T_c$  & $0.772$ & $0.594$ & $1.228$ \\ \botrule
\end{tabular}}
\label{tab:table1}
\end{table}

\section{The social network}
\label{sec:evolsocnet}

The affinity matrix drives the interaction via the selection mechanism. It hence
can be interpreted as the {\em adjacency} matrix of the
underlying {\em social network}, i.e. the network of social ties that
influences 
the exchange of opinions between acquaintances, as mediated by the encounters. Because the
affinity is a 
dynamical variable of the model, we are actually focusing on 
 an {\em adaptive} social network~\cite{GrossBlasius,ZESM} : The network
 topology influences in turn the dynamics of opinions, this latter providing a feedback on
 the network itself and so modifying its topology. In other words, the evolution 
of the topology is inherent
 to the dynamics of the model because of the  proposed self-consistent
 formulation and not imposed a priori as an additional, external ingredient,
 (as e.g. rewire and/or add/remove links according to a given
 probability~\cite{HN,KB} once the state variables have been updated).

 \begin{remark}[Weighted network]
   Let us observe that the affinity assumes positive real values, hence we can
   consider 
a weighted social networks, where agents weigh the relationships. Alternatively, one can 
introduce a cut-off parameter, $\alpha_f$: agents $i$ and $j$ are 
socially linked if and only if the recorded relative affinity is large enough, meaning
 $\alpha_{ij}>\alpha_f$. Roughly, the agent chooses its
closest friends among all his neighbors.

The first approach avoids the introduction of non--smooth functions and it is 
suitable to carry on the analytical calculations. The latter results more straightforward 
for numerical oriented applications.
 \end{remark}

As anticipated, we are thus interested in analyzing the model, for a specific choice of the parameters, $\alpha_c$, 
$\sigma$ and $\Delta O_c$, yielding to consensus, and studying the evolution
of the network topology, here analyzed via standard network indicators: the average value
 of {\em weighted degree}, the {\em cluster coefficient} and the {\em averaged
  shortest path}. These quantities will be  quantified for (i)
a fixed population, monitoring their time dependence; (ii) as a function of the
population size, photographing the dynamics at convergence, namely when consensus has been reached.

\subsection{Time evolution of weighted degree}
\label{ssec:degree}

The simplest and the most intensively studied one--vertex (i.e. local)
characteristic is the node {\em degree}~\footnote{Let us observe that the
  affinity may not be symmetric and thus the inspected social network will be directed.
  One has thus to distinguish between {\em In--degree}, $k_{in}$, being the
  number of {\em incoming edges} of a vertex and {\em Out--degree}, $k_{out}$,
  being the number of its {\em outgoing edges}. In the following we will be
interested only in the outgoing degree, from here on simply referred to as to degree.}: 
the {\em total number of its
  connections} or its nearest neighbors. Because we are dealing with a weighted
network we can also introduce the {\em weighted node degree}, also 
called {\em node strength}~\cite{BBPV2004}, namely
$s_i(t)=\sum_{j}\alpha_{ij}^t/(N-1)$. Its mean value averaged over the whole
network reads: 
\begin{equation}
  \label{eq:strength}
  <s>(t) = \frac{1}{N}\sum_{i=1}^N s_i(t)\, .
\end{equation}
Let us observe that the normalization factor 
$N-1$ holds for a population of
$N$ agents, self-interaction being disregarded, $<s>$ belongs hence to the interval $[0,1]$ and having eliminated the relic of the 
population size, one can properly compare quantities calculated for networks made of different number of agents.

All these quantities evolve in time because of the dynamics of the opinions
and/or affinities. Passing to continuous time and
using the second relation of~\eqref{eq:themodel}, we obtain: 
\begin{equation}
  \label{eq:meandegevolv}
  \frac{d}{dt}<s> = \frac{1}{N(N-1)}\sum_{i,j=1}^N \frac{d}{dt}\alpha_{ij}^t\, .
\end{equation}
Let us observe that the evolution of affinity and opinion can be
decoupled when $\Delta O_c=1$. For $\Delta O_c <1$, this is not formally
true. However on can argue for an 
 approximated strategy~\cite{opi3}, by replacing the step function $\Gamma_2$ by its time
average counterpart $\gamma_2$, where the dependence in $\Delta O^t_{ij}$ has been silenced. 
In this way, we obtain form~\eqref{eq:meandegevolv}
\begin{equation}
  \label{eq:meanalphaevolv}
  \frac{d}{dt}<s> =\frac{\gamma_2}{N(N-1)}\sum_{i,j=1}^N \alpha^t_{ij}(1 -
\alpha^t_{ij})=\gamma_2 (<s>-<s^2>)\, ,
\end{equation}
where $<s^2> = \sum \alpha_{ij}^2/(N(N-1))$. Let us observe that $\gamma_2$ is
of the order of $1/N^2$ times, a factor taking care of the asynchronous
dynamics~\cite{opi3}.

\begin{figure}[th]
\centerline{\psfig{file=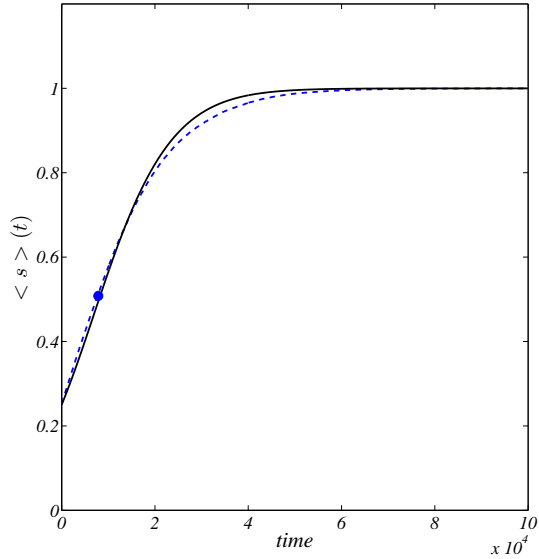,width=8cm}}
\vspace*{8pt}
\caption{Evolution of $<s>(t)$. Dashed line (blue on-line) refers to
  numerical simulations with parameters $\alpha_c=0.5$, $\Delta O_c=0.5$
  and $\sigma=0.3$. The full line (black on-line) is the analytical
  solution~\eqref{eq:sanal} with a best fitted parameter $\gamma_2=1.6 \,
  10^{-4}$. The dot denotes the convergence time in the opinion space 
to the consensus state, for the used parameters affinities did not yet converge.
 Let us observe in fact that affinities and opinions do converge
 on different time scale~\cite{opi3}.}  
\label{fig:alphamed}
\end{figure}

In~\cite{evoden} authors proved that~\eqref{eq:meanalphaevolv} can be
analytically solved once we provide the initial distribution of node strengths
(see~\ref{app:analsol} for a short discussion of the involved 
methods). Assuming $s_i(0)$ to be uniformly distributed in $[0,1/2]$, we get
the following solution (see Fig.~\ref{fig:alphamed}):
\begin{equation}
  \label{eq:sanal}
  <s>(t)=\frac{e^{\gamma_2 t}}{e^{\gamma_2 t}-1}-\frac{2e^{\gamma_2
      t}}{(e^{\gamma_2 t}-1)^2}\log\left(\frac{e^{\gamma_2 
          t}+1}{2}\right)\, ,
\end{equation}

Using similar ideas we can prove~\cite{evoden} that the variance
$\sigma^2_{s}(t)=<s^2>-<s>^2$ is analytically given by
\begin{equation}
  \label{eq:sigmaUD}
  \sigma^2(t)=\frac{2e^{2\gamma_2 t}}{(e^{\gamma_2 t}-1)^2(e^{\gamma_2
      t}+1)}-\frac{4e^{2\gamma_2 t}}{(e^{\gamma_2
      t}-1)^4}\left[\log\left(\frac{e^{\gamma_2 
          t}+1}{2}\right)\right]^2\, .
\end{equation}

The comparison between analytical and numerical profiles is enclosed in Fig.~\ref{fig:alphamed}, where the evolution of $<s>(t)$ is traced. Let us observe that here $\gamma_2$ will serve as a fitting parameter, when testing the adequacy of the proposed 
analytical curves versus direct simulations, instead of using its computed numerical value~\cite{opi3}. 
The qualitative correspondence is rather satisfying, so confirming the correctness of the analytical results reported above.

\begin{figure}[th]
\centerline{\psfig{file=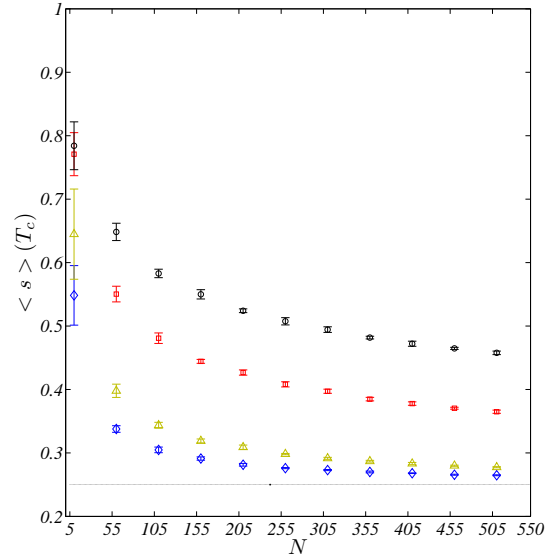,width=8cm}}
\vspace*{8pt}
\caption{Average node strength at convergence as a function of the population
  size. Parameters are $\Delta O_c=0.5$, $\sigma=0.5$ and four values of
  $\alpha_c$ have been used : ($\Diamond$) $\alpha_c=0$, ($\triangle$)
  $\alpha_c=0.25$, ($\Box$) $\alpha_c=0.5$ and ($\bigcirc$)
  $\alpha_c=0.75$. Vertical bars are standard deviations computed over $10$
  replicas of the numerical simulation using the same initial conditions.} 
\label{fig:Ngradomedio}
\end{figure}

Assume $T_c$ to label the time needed for the consensus to be reached. Clearly,  $T_c$ depends on the size of the
simulated system~\footnote{In~\cite{pre,epl} it was shown that $T_c$ scales   
faster than linearly but slower than 
quadratically with respect to the population size $N$.}. From the above relation  (\ref{eq:sanal}),
the average node strength at convergence as an implicit function of the population size 
$N$ reads:
\begin{equation}
  \label{eq:sTc}
  <s>(T_c(N))=\frac{e^{\gamma_2(N) T_c(N)}}{e^{\gamma_2(N)
      T_c(N)}-1}-\frac{2e^{\gamma_2(N) T_c(N)}}{(e^{\gamma_2(N)
      T_c(N)}-1)^2}\log\left(\frac{e^{\gamma_2(N) T_c(N)}+1}{2}\right)\, , 
\end{equation}
where we emphasized the dependence of $\gamma_2$ and $T_c$ on $N$. However,
as already observed $\gamma_2(N)=\mathcal{O}\left(N^{-2}\right)$ and $T_c(N)=
\mathcal{O}\left(N^{a}\right)$, with $a\in (1,2)$. Hence
$\gamma_2(N)T_c(N)\rightarrow~0$ when $N\rightarrow \infty$ and thus 
$<s>(T_c(N))$ is predicted to be a decreasing function of the population size $N$, which
converges to the asymptotic value $1/4$, a value identical to the initial average node
strength (see Fig.~\ref{fig:Ngradomedio}), given the selected initial condition. In
sociological terms this means that even when consensus is achieved the larger
the group the smaller, on average, the number of local acquaintances. This is a second 
conclusion that one can reach on the basis of the above analytical developments. 

\subsection{Small world}
\label{sec:smallworld}

Several social networks exhibit the remarkable property that one can reach an arbitrary far member of the community,
via a relatively small number of intermediate acquaintances. This holds true irrespectively of the size of the underlying network.
Experiments~\cite{milgram} have been devised to quantify the
\lq\lq degree of separation\rq\rq in real system, 
and such phenomenon is nowadays termed the \lq\lq small world\rq\rq effect, also referred
to as the \lq\lq six degree of separation\rq\rq. 

On the other hand several, models have been proposed~\cite{WS1998,NW1999} to construct complex
networks with the 
small world property. Mathematically, one requires that the average
shortest path grows at most logarithmic with respect to the network size, while the network still displays a
large  clustering coefficient. Namely, the network has an average shortest path
comparable 
to that of a random network, with the same number of nodes and links, while
the clustering coefficient is instead significantly larger. 

In this section we present numerical results aimed at describing the
time evolution of both the average shortest path and the clustering coefficient of
the social network emerging from the model. As before, the parameters are set so to induce the convergence to a 
consensus state in the opinion space.



We will be particularly interested in their asymptotic solutions, terming the associated values respectively 
$<\ell>(T_c)$ and $C(T_c)$ once the consensus state has been achieved. 

\begin{figure}[th]
\centerline{\psfig{file=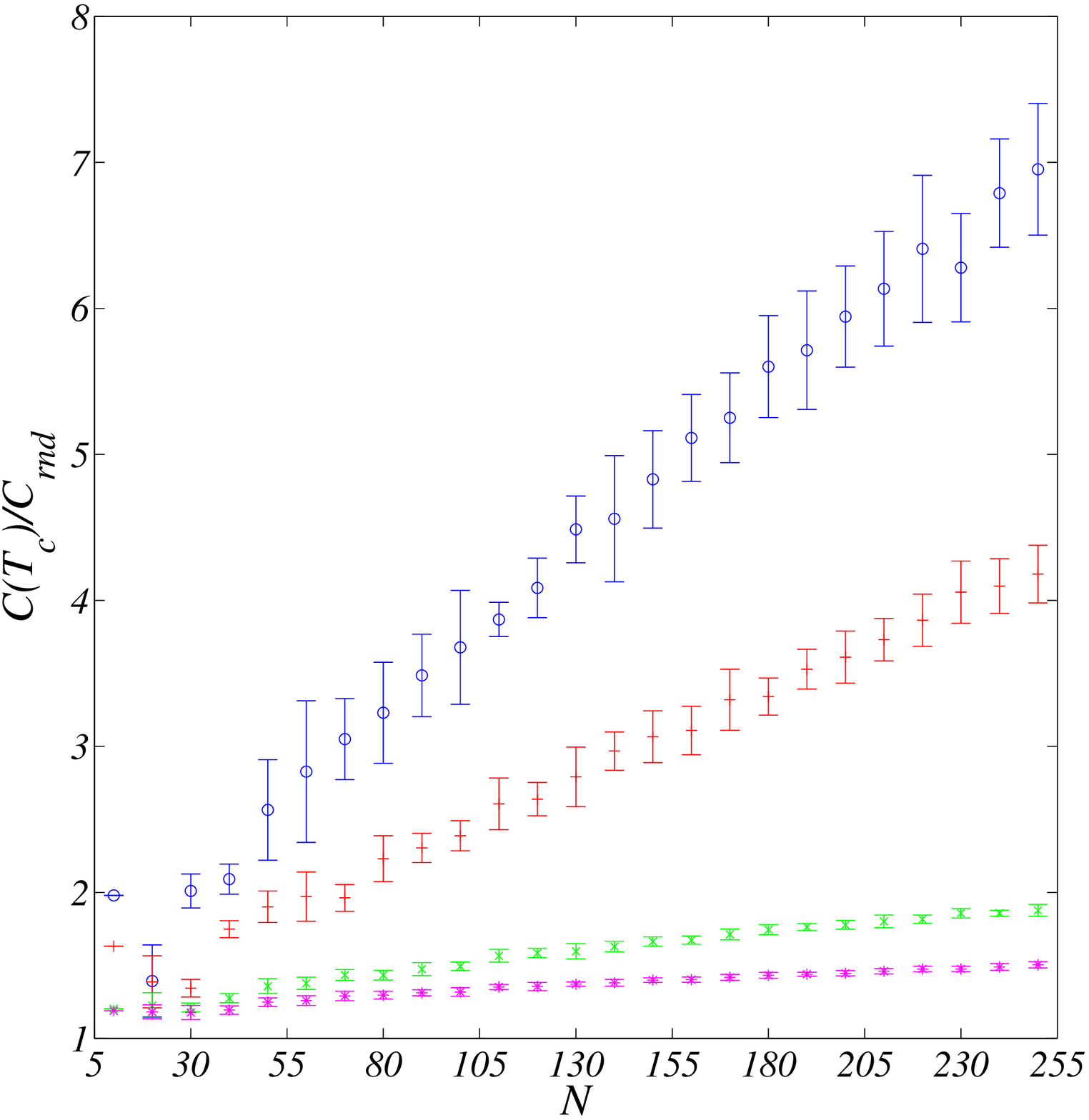,width=7cm}
\quad \psfig{file=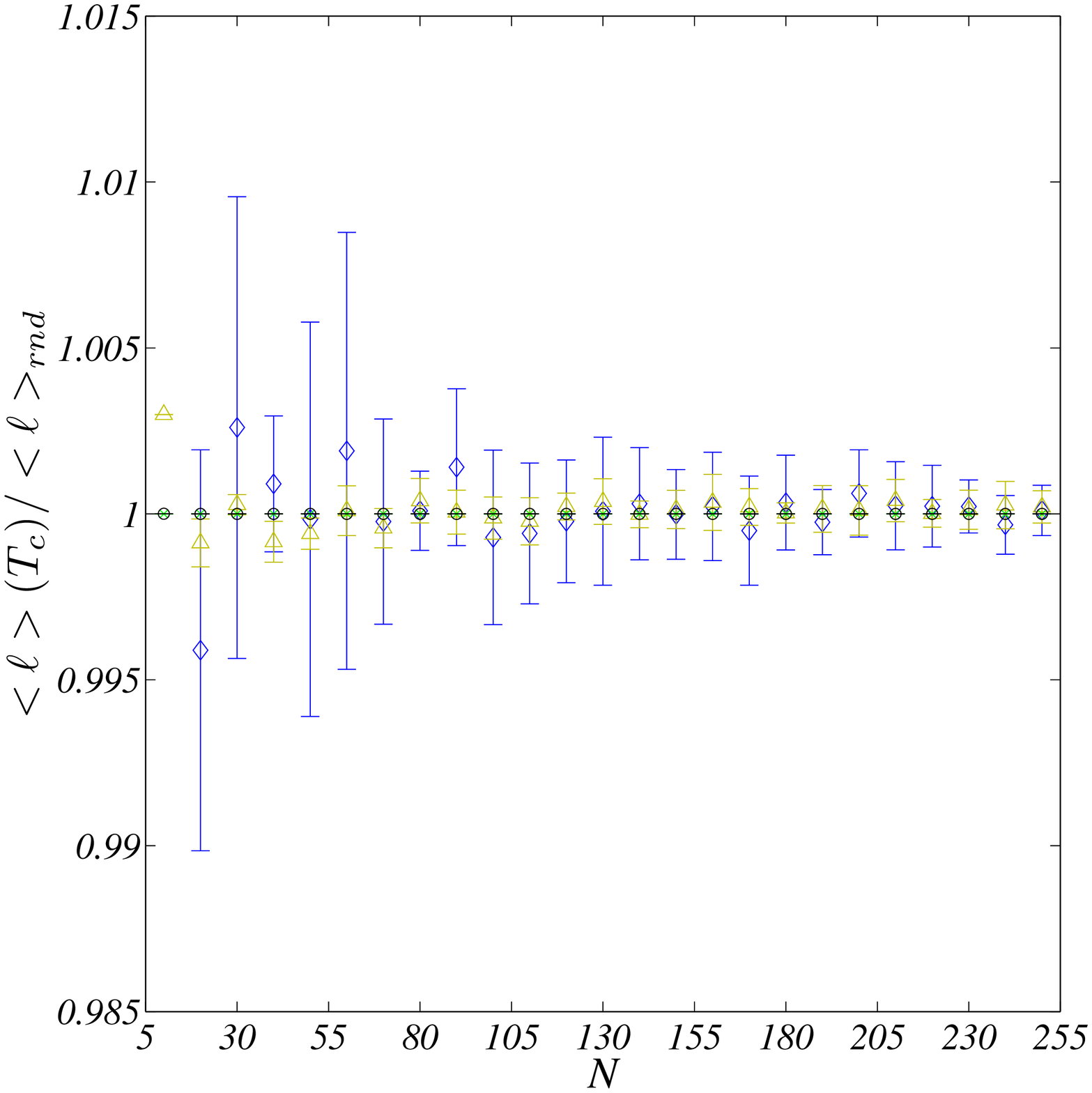,width=7cm}}
\vspace*{8pt}
\caption{Normalized clustering coefficient (left panel) and normalized
  average mean path (right panel) as functions of the network size at the convergence
  time. Parameters 
  are $\Delta O_c=0.5$, $\sigma=0.5$ and four values of 
  $\alpha_c$ have been used : ($\Diamond$) $\alpha_c=0$, ($\triangle$)
  $\alpha_c=0.25$, ($\Box$) $\alpha_c=0.5$ and ($\bigcirc$)
  $\alpha_c=0.75$. Vertical bars are standard deviations computed over $10$
  repetitions.}
\label{fig:NevolCCrndEll}
\end{figure}

In Fig.~\ref{fig:NevolCCrndEll} we report these quantities 
(normalized to the homologous values estimated for a random network with identical 
number of nodes and links) versus 
the system size. The (normalized) clustering coefficient is sensibly larger than one, this effect being more pronounced the
smaller the value of $\alpha_c$. On the other hand the (normalized) average
shortest path is always very close to $1$. 

Based on the above we are hence brought to conclude that the
social network emerging from the opinion exchanges, has the small world
property.  This is a remarkable feature because the social
network evolves guided by the opinions, as it does in reality, and not result from 
an artificially imposed recipe. 

\subsection{Weak ties}
\label{ssec:weakties}

Social networks are characterized by the presence of hierarchies of well tied
small  groups of acquaintances, that are possibly linked to other such groups via
\lq\lq weak ties\rq\rq. According to 
Granovetter~\cite{Granovetter1983} these weak links are fundamental for the
cohesion of the society, being at the basis of the social tissue, so motivating the
statement \lq\lq the strength of weak ties\rq\rq.

The smallest group in a social network is composed by three individuals
sharing high mutual affinities, in term of network theory they form a {\em
  clique}~\cite{AB2002}, i.e. a maximal complete graph composed by three
nodes. This can of 
course 
be generalized to larger maximal complete graphs, defining thus $m$-cliques. 

The
degree of cliqueness of a social network is hence a measure of its
cohesion/fragmentation: the presence of a large number of $m$-cliques together
with very few, $m^{\prime}$-cliques, for $m^{\prime}>m$, means
that the population is actually fragmented into small pieces, of size $m$ not
strongly interacting each other.

We are  interested in studying such phenomenon within the social network
emerging from the opinion dynamics model here considered, still operating in 
consensus regime. To this end we proceed as follows. We introduce a
cut--off parameter $\alpha_f$ used to binarize the affinity matrix, which
hence transforms into a 
an adjacency matrix $a$. More precisely, agents $i$ and $j$ will be
connected, i.e. $a_{ij}=1$, if and only if $\alpha_{ij}\geq \alpha_f$. Once 
the adjacency matrix is being constructed, we compute the number of $m$--cliques in the
network. Let us observe that this last step is highly time consuming, being the
clique problem NP-complete. We thus restrict our analysis to the cases $m\in\{ 3,4,5 \}$.

\begin{figure}[th]
\centerline{\psfig{file=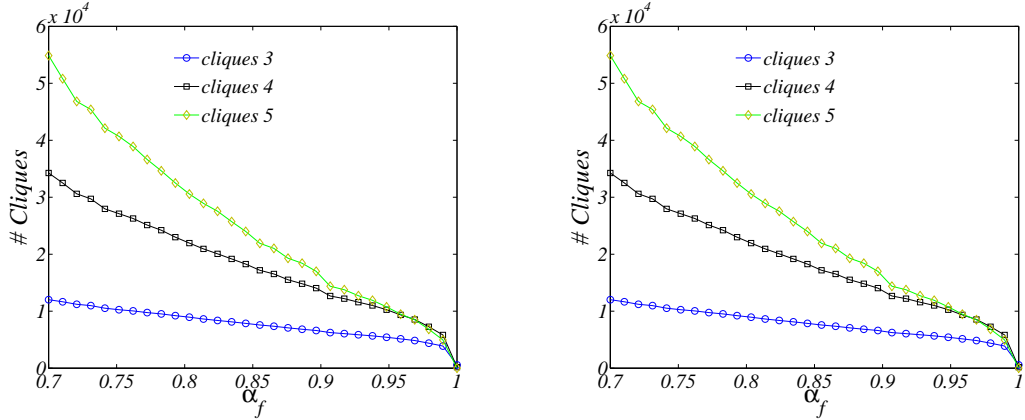,width=7cm}
\quad \psfig{file=Simulaz20080626_003.eps,width=7cm}}
\vspace*{8pt}
\caption{Number of $3$, $4$ and $5$--cliques in the social network once
  consensus has been achieved. Parameters are $N=100$, $\Delta O_c=0.5$,
  $\alpha_c=0.5$. Right panel, $\sigma=0.5$, the network exhibits the weak ties
  property. Left panel, $\sigma=0.1$, the network does not display the weak ties phenomenon.}
\label{fig:cliques}
\end{figure}

For small values of $\alpha_f$ the network is almost complete, while for
large ones it can in principle fragment into a vast number of finite small groups
of agents. As reported in the inset of right panel of
Fig.~\ref{fig:cliques}, for $\alpha_f\sim 1$ only $3$--cliques are present. 
Their number rapidly increases as $\alpha_f$ is lowered. On the other hand
for $\alpha_f\sim 0.98$ few $4$--cliques emerge while $5$--cliques appear
around $\alpha_f\sim 0.73$. This means that the social networks is mainly
composed by $3$--cliques, i.e. agents sharing high mutual affinities, that are 
connected together to form larger cliques,
for instance $4$ and $5$--cliques by weaker links, i.e. whose mutual affinities  are lower than the above ones.

Results reported in left panel of Fig.~\ref{fig:cliques} show that for specific
parameter values, still falling into the class deputed to the consensus dynamics, the model does
not present the weak ties phenomenon:  $3$, $4$ and $5$-cliques are all
present at the same time for large values of $\alpha_f$. This is an important point that will
deserve future investigations. Let us observe here that the observed differences 
 stem  from the social temperature. 
 

\section{Conclusion}
\label{sec:conc}

Social system and opinion dynamics models are intensively investigated within simplified mathematical schemes. 
One of such model is here revisited and analyzed. The evolution of the underlying network of connections, here emblematized 
by the mutual affinity score, is in particular studied. This is a dynamical quantity which adjusts all along the system evolution, as follows a 
complex coupling with the opinion variables.  In other words, the embedding social structure is adaptively created and not a 
priori assigned, as it is customarily done.  Starting from this setting, the model is solved analytically, under specific approximations. The
functional dependence on time of the networks mean characteristics are consequently elucidated.   The obtained solutions correlate
with direct simulations, returning a satisfying agreement. Moreover, the structure of the social network is numerically monitored,
via a set of classical indicators. Small world effects, as well weak ties connections, are found as an emerging property of the model. It 
is remarkable that such properties, ubiquitous in nature, are spontaneously generated within a simple scenario which accounts for a minimal number 
of ingredients, in the context of a genuine self-consistent formulation. 

\appendix

\section{On the momenta evolution}
\label{app:analsol}
 
The aim of this section if to present and sketch the proof of the result used
to study the evolution of the momenta of the affinity distribution. We refer
the interested reader to~\cite{evoden} where a more detailed analysis is
presented in a general setting.

For the sake of simplicity, let us label the $N(N-1)$ affinities values
$\alpha_{ij}$ by  $x_l$, upon assigning a specific re-ordering of the entries.
Hence $\vec{x}$ is a vector with $M=N(N-1)$
elements. As previously recalled~(\ref{eq:meanalphaevolv}), we assume each $x_l$ to
obey a first order differential equation of the logistic type, once time has been rescaled, namely: 
\begin{equation}
  \label{eq:Ver1}
  \frac{d x_l}{d t}= x_l (1-x_l)\, .
\end{equation}
The initial conditions will be denoted as $x_l^0$.  

Let us observe that each component $x_l$ evolves independently from the
other. We can hence imagine to deal with $M$ replicas of a process ruled by
by~\eqref{eq:Ver1} whose initial conditions are distributed according to some
given function. We are interested in computing the momenta of the $x$ variable
as functions of the initial distribution. The $m$-th momentum is given by:
\begin{eqnarray}
  \label{eq:mmom2}
  <x^{m}>(t)=\frac{\left(x_1(t)\right)^m+\dots+\left(x_M(t)\right)^m}{M}\, ,
\end{eqnarray}
and its time evolution is straightforwardly obtained deriving~(\ref{eq:mmom2}) 
and making use of Eq.~(\ref{eq:Ver1}):
\begin{eqnarray}
  \label{eq:xmom}
  \frac{d}{dt}<x^m>(t)&=&\frac{1}{M}\sum_{i=1}^M
  \frac{dx_l^m}{dt}=\frac{m}{M}\sum_{l=1}^N x_l^{m-1} \frac{dx_l}{dt}\nonumber
  \\
&=&\frac{m}{M}\sum_{l=1}^N x_l^{m-1}
x_l(1-x_l)=m\left(<x^m>-<x^{m+1}>\right)\, .
\end{eqnarray}

To solve this equation we introduce the {\it time dependent moment
  generating function}, $G(\xi,t)$, 
\begin{equation}
  \label{eq:formF}
  G(\xi,t):=\sum_{m=1}^{\infty}\xi^m <x^m>(t)\, .
\end{equation}
This is a formal power series whose Taylor coefficients are the momenta
of the distribution that we are willing to reconstruct, task that can 
be accomplished using the following relation: 
\begin{equation}
  \label{eq:momen}
  <x^m>(t):=\frac{1}{m!}\frac{\partial^m G}{\partial
    \xi^m}\Big |_{\xi=0}\, . 
\end{equation}

By exploiting the evolution's law for each $x_l$, 
we shall here obtain a partial differential equation governing
  the behavior of $G$. Knowing $G$  
will eventually enables us to calculate any sought momentum via multiple
differentiation with respect to $\xi$ as stated in~(\ref{eq:momen}).

On the other hand, 
by differentiating~(\ref{eq:formF}) with respect to
time, one obtains : 
\begin{equation}
  \label{eq:diff1}
  \frac{\partial G}{\partial t}=\sum_{m\geq 1}\xi^m\frac{d<
    x^m>}{dt}=\sum_{m\geq 1}m\xi^m\left(<x^m>-<x^{m+1}>\right)\, , 
\end{equation}
where used has been made of Eq.~(\ref{eq:xmom}). We can now re-order the terms
so to express the right hand side as a function of $G$~\footnote{Here 
 the following algebraic relations are being used:
\begin{equation}
  \label{eq:derx}
  \xi\partial_{\xi} G(\xi,t)=\xi\sum_{m\geq 1}m\xi^{m-1}<x^m> = \sum_{m\geq
    1}m\xi^{m}<x^m> \, , \nonumber
\end{equation}
and 
\begin{eqnarray}
  \label{eq:der2x}
  \xi\partial_{\xi} \frac{G(\xi,t)}{\xi}&=&\xi\partial_{\xi}\sum_{m\geq
    1}\xi^{m-1}<x^m> = 
  \xi\sum_{m\geq 1}(m-1)\xi^{m-2}<x^m>\nonumber \\ &=&\sum_{m\geq
    1}(m-1)\xi^{m-1}<x^m> \, \nonumber
\end{eqnarray}
Renaming the summation index, $m-1\rightarrow m$, one finally gets (note
the sum still begins with $m=1$):
\begin{equation}
  \label{eq:derxx}
  \xi\partial_{\xi} \frac{G(\xi,t)}{\xi}=\sum_{m\geq 1}m\xi^{m}<x^{m+1}>
  \,\nonumber.
\end{equation}}
and finally obtain the following non--homogeneous linear partial differential
equation:  
\begin{equation}
    \label{eq:forF}
  \partial_t G - (\xi-1)\partial_{\xi} G =\frac{G}{\xi}\, .
\end{equation}

Such an equation can be solved for $\xi$ close to zero (as in the end of the
procedure  
we shall be interested in evaluating the derivatives at 
$\xi=0$, see Eq. (\ref{eq:momen})
) and for all positive $t$. To this end we shall specify the 
initial datum: 
\begin{equation}
  \label{eq:inidat}
  G(\xi,0)=\sum_{m\geq 1}\xi^m<x^m>(0)=\Phi(\xi)\, ,
\end{equation}
i.e. the initial momenta or their distribution.

Before turning to solve~(\ref{eq:forF}), we first simplify it by introducing
\begin{equation} 
  \label{eq:fF}
  G=e^g \quad {\rm namely}\quad g=\log G\, ,
\end{equation}
then for any derivative we have $\partial_* G = G\partial_*g$, where $*=\xi$
or $*=t$, thus~(\ref{eq:forF}) is equivalent to 
\begin{equation}
  \label{eq:forf}
  \partial_t g-(\xi-1)\partial_{\xi} g =\frac{1}{\xi}\, ,
\end{equation}
with the initial datum
\begin{equation}
  \label{eq:initdatf}
  g(\xi,0)=\phi(\xi)\equiv \log \Phi(\xi)\, .
\end{equation}

This latter equation can be solved using the {\it method of the
  characteristics}, here represented by:
\begin{equation}
  \label{eq:char}
  \frac{d\xi}{dt}=-(\xi-1)\, ,
\end{equation}
which are explicitly integrated to give:
\begin{equation}
  \label{eq:charsol}
  \xi(t)=1+(\xi(0)-1)e^{-t}\, ,
\end{equation}
where $\xi(0)$ denotes $\xi(t)$ at $t=0$. Then the function
$u(\xi(t),t)$ defined by: 
\begin{equation}
  \label{eq:funcu}
  u(\xi(t),t):=\phi(\xi(0))+\int_0^t\frac{1}{1+(\xi(0)-1)e^{-s}}\, ds\, , 
\end{equation}
is the solution of~(\ref{eq:forf}), restricted to the
characteristics. Observe that $u(\xi(0),0)=\phi(\xi(0))$, so~(\ref{eq:funcu})
solves also the initial value problem. 

Finally the solution of~(\ref{eq:initdatf}) is obtained from $u$ by reversing
the relation between $\xi(t)$ and $\xi(0)$, i.e. $\xi(0)=(\xi(t)-1)e^t+1$:
\begin{equation}
  \label{eq:soluf}
  g(\xi,t)=\phi\left((\xi-1)e^t+1\right)+\lambda(\xi,t)\, ,
\end{equation}
where $\lambda(\xi,t)$ is the value of the integral in the right hand side
of~(\ref{eq:funcu}). 

This integral can be straightforwardly computed as follows (use the
change of variable $z=e^{-s}$):
\begin{equation}
\label{eq:int1}
  \lambda = \int_0^t\frac{1}{1+(\xi(0)-1)e^{-s}}\, ds
  =\int_1^{e^{-t}}\frac{-dz}{z}\frac{1}{1+(\xi(0)-1)z}\, , 
\end{equation}
which implies 
\begin{eqnarray}
  \label{eq:int2}
  \lambda &=&
  -\int_1^{e^{-t}}dz\left(\frac{1}{z}-\frac{\xi(0)-1}{1+(\xi(0)-1)z}\right)= -\log 
  z+\log (1+(\xi(0)-1)z)\Big |_{1}^{e^{-t}}\nonumber \\ &=&t+\log
  (1+(\xi(0)-1)e^{-t})-\log \xi(0)\, . 
\end{eqnarray}

According to~(\ref{eq:soluf}) the solution $g$ is then
\begin{equation}
  \label{eq:soluf2}
  g(\xi,t)=\phi\left((\xi-1)e^t+1\right)+t+\log \xi -\log ((\xi-1)e^{t}+1)\, ,
\end{equation}
from which $G$ straightforwardly follows: 
\begin{equation}
  \label{eq:solFF}
  G(\xi,t)=\Phi\left((\xi-1)e^{t}+1\right)\frac{\xi e^t}{(\xi-1)e^{t}+1}\, .
\end{equation}

As anticipated, the function $G$ makes it possible to estimate any momentum  (\ref{eq:momen}). As an 
example, the mean value correspond to setting  $m=1$, reads: 
\begin{eqnarray}
  \label{eq:mom1f}
  <x>(t)&=&\partial_{\xi}G\Big |_{\xi=0}=\Big[
  \Phi^{\prime}\left(1+(\xi-1)e^{t}\right) e^{t}
  \frac{\xi e^t}{(\xi-1)e^{t}+1}\nonumber\\ &+&\Phi\left(1+(\xi-1)e^{t}\right)
  e^{t}\frac{(\xi-1)e^{t}+1-\xi
    e^{t}}{\left(1+(\xi-1)e^{t}\right)^2}\Big]\Big |_{\xi=0}\nonumber\\
  &=&\frac{e^{t}}{1-e^{t}}\Phi(1-e^{t})\, .   
\end{eqnarray}

In the following section we shall turn to considering a specific application 
in the case of uniformly distributed values of affinities.

\subsection{Uniform distributed initial conditions}

The initial data $x_l^0$ are assumed to span uniformly the bound interval
$[0,0.5]$, thus the probability distribution $\psi(x)$ clearly reads~\footnote{We hereby assume to
  sample over a large collection of independent replica of the system under
  scrutiny 
($M$ is large). Under this hypothesis one can safely adopt a continuous
approximation for the distribution of allowed initial data. Conversely, if the
number of  
realizations is small, finite size corrections need to be
included~\cite{evoden}.}:  
\begin{equation}
  \psi(x)=  \left\{ \begin{array}{ll}
 2 & \textrm{ if $x\in[0,1/2]$}\\
 0  & \textrm{ otherwise}
 \end{array} \right. \, ,
\end{equation}
and consequently the initial momenta are:
\begin{equation}
  \label{eq:mommexUD}
  <x^m>(0)=\int_0^1 \xi^m
  \psi(\xi) d\xi =\int_0^{1/2}2\xi^m\, d\xi = \frac{1}{m+1}\frac{1}{2^m}\, .
\end{equation}

Hence the function $\Phi$ as defined in~(\ref{eq:inidat}) takes the form:
\begin{equation}
  \label{eq:phiUD}
  \Phi(\xi)=\sum_{m\geq 1}\frac{1}{m+1}\frac{\xi^m}{2^m}\, .
\end{equation}
A straightforward algebraic manipulation allows us to
re-write~(\ref{eq:phiUD}) as follows:
\begin{equation}
  \sum_{m\geq 1}\frac{y^m}{m+1}=\frac{1}{y}\int_0^y \sum_{m\geq 1}z^m\,
  dz=\frac{1}{y}\int_0^y \frac{z}{1-z}\, dz = -1-\frac{1}{y}\log (1-y)\, ,
\end{equation}
thus
\begin{equation}
  \label{eq:phiUDfin}
  \Phi(\xi)=-1-\frac{2}{\xi}\log \left(1-\frac{\xi}{2}\right)\, .
\end{equation}

We can now compute the time dependent moment generating function, $G(\xi,t)$,
given by~(\ref{eq:solFF}) as:
\begin{equation}
  \label{eq:FUD}
  G(\xi,t)=\frac{\xi e^{t}}{(\xi -1)e^{t}+1}\left[-1-\frac{2}{(\xi -1)e^{t}+1}\log \left(
      1-\frac{(\xi -1)e^{t}+1}{2}\right)\right]\, ,
\end{equation}
and thus recalling~(\ref{eq:momen}) we get
\begin{eqnarray}
  \label{eq:ameda2t}
  <x>(t)&=&\frac{e^{t}}{e^{t}-1}-\frac{2e^{t}}{(e^{t}-1)^2}\log\left(\frac{e^{t}+1}{2}\right)\\ 
<x^2>(t)&=&\frac{e^{2 t}}{(e^{t}-1)^2}+\frac{4e^{2
    t}}{(e^{t}-1)^3}\log\left(\frac{e^{t}+1}{2}\right)+\frac{2e^{2t}}{(e^{t}-1)^2(e^{ t}+1)}\nonumber\, .
\end{eqnarray}
Let us observe that $<x>(t)$ deviates from the logistic growth to which all
the single variable $x_i(t)$ does obey.

For large enough times, the distribution of the variable outputs
is in fact concentrated around the asymptotic value $1$ with an associated
variance (calculated from the above momenta) which decreases monotonously with
time. 

Let us observe that a naive approach would suggest interpolating the averaged
numerical profile with a solution of the  
logistic model whose initial datum $\hat{x}^0$ acts as a free
  parameter to be adjusted to its best fitted value: as it is
  proven in~\cite{evoden} this procedure yields a significant
discrepancy, which could be possibly misinterpreted as a failure of the
underlying   
logistic evolution law. For this reason, and to avoid drawing erroneous
conclusions when ensemble averages are computed,   
attention has to be payed on the role of initial conditions.

\bibliographystyle{ws-acs}
\bibliography{CarlettiFanelliRighi}

\end{document}